\documentclass[reprint,aps,prd,onecolumn,notitlepage,showpacs,nofootinbib,preprintnumbers,superscriptaddress]{revtex4-1}
\usepackage{amsmath}
\usepackage{amsfonts}
\usepackage{amssymb}
\usepackage{latexsym}
\usepackage{enumerate}
\usepackage{color,xcolor}
\usepackage{graphicx}
\usepackage{bm}
\usepackage{epsfig}
\usepackage[english]{babel}
\usepackage{hyperref}
\usepackage{times}
\usepackage{comment}
\usepackage{epstopdf}

\def\dom{\left(d\theta^2+\sin^2\theta d\phi^2\right)}
\def\rmin{r_\mathrm{min}}
\def\rin{r_\mathrm{in}}
\def\rh{r_\mathrm{H}}
\def\rph{r_\mathrm{ph}}
\def\ro{r_\mathrm{o}}
\def\risco{r_\mathrm{ISCO}}
\def\rsh{R_\mathrm{sh}}
\def\ash{\alpha_\mathrm{sh}}
\def\bph{b_\mathrm{ph}}
\def\dl{dl_{\mathrm{prop}}}
\def\agr{\theta_{\mathrm{g}}}
\def\muas{\mu\mathrm{as}}

\begin{document}

\title{Images of nonsingular nonrotating black holes in conformal gravity}
\author{Zhi-Shuo Qu}
\author{Towe Wang}
\email[Electronic address: ]{twang@phy.ecnu.edu.cn}
\affiliation{Department of Physics, East China Normal University, Shanghai 200241, China\\}
\author{Chao-Jun Feng}
\email[Electronic address: ]{fengcj@shnu.edu.cn}
\affiliation{Division of Mathematical and Theoretical Physics, Shanghai Normal University, Shanghai 200234, P.R.China\\ \vspace{0.2cm}}
\date{\today\\ \vspace{1cm}}
\begin{abstract}
The accretion disk around a black hole and its emissions play an essential role in theoretical analysis of the black hole image. In the literature, two analytical toy models of accretions are widely adopted: the spherical model and the thin disk model. They are different geometrically but both thin optically. We polish them for free-falling accretions around static spherical black holes. As an application, we investigate the images of a class of nonsingular black holes conformally related to the Schwarzschild black hole. These black holes are vacuum solutions of a family of conformal gravity theories. Results are compared with the Schwarzschild black hole of the same mass. Our results indicate that the conformal factor does not affect the shadow radius seen by distant observers, but it leaves an imprint on the intensity image of black hole.
\end{abstract}


\maketitle




\section{Introduction}\label{sect-intro}
The shadow cast is a prominent feature of black hole (BH) \cite{Synge:1966okc,Bardeen:1973,Luminet:1979nyg}. It is more accessible to observation when the BH is surrounded by an optically thin and geometrically thick emission region \cite{Falcke:1999pj}, and success has been achieved by the Event Horizon Telescope firstly in the center of galaxy M87 \cite{EventHorizonTelescope:2019dse} and recently in the center of the Milky Way \cite{EventHorizonTelescope:2022xnr}. The success in observation opened a new window to probe the geometry of spacetime near BHs. Furthermore, since the geometry of vacuum BHs in 4-dimensional Einstein theory is subject to uniqueness theorems \cite{Israel:1967wq,Robinson:1975bv}, one can test Einstein theory by confronting predictions of the Schwarzschild or Kerr solution against the observational data of shadow \cite{Psaltis:2018xkc,Bambi:2019tjh,Vagnozzi:2019apd,Vagnozzi:2019apd,Allahyari:2019jqz,Khodadi:2020jij,Ghosh:2022gka}. For recent reviews and progresses on BH shadows, the readers are referred to Refs. \cite{Perlick:2021aok,Bronzwaer:2021lzo,Wang:2022kvg,Vagnozzi:2022moj,Chen:2022scf} and references therein.

In vacuum Einstein theory with a 4-dimensional spacetime, the Schwarzschild BH is the unique static asymptotically-flat spacetime exterior to a regular event horizon \cite{Israel:1967wq}. In non-vacuum or non-Einstein theories, there are other BHs or ultra-compact stars that can imitate the shadow of Schwarzschild BH, usually with different masses. Then it is intriguing to ask who is the best imitator in this imitation game, or specifically, whether the Schwarzschild shadow can be mimicked by an object of the same mass. To the best of our knowledge, the vector boson star is the only studied object with this ability in the literature. According to Refs. \cite{Olivares:2018abq,Herdeiro:2021lwl}, when masses are equal, the radius of maximal angular velocity of the vector boson star can be the same as the radius of the innermost stable circular obit (ISCO) of Schwarzschild BH, giving rise to the same shadow size.

The present paper is devoted to another example, the nonsingular nonrotating BH in conformal gravity, which is related to the Schwarzschild solution by a conformal transformation \cite{Modesto:2016max,Bambi:2016wdn,Bambi:2017yoz}. For brevity, we will refer to it as the conformal Schwarzschild BH. As we will show, for distant observers, the conformal and the standard Schwarzschild BHs of the same mass have the same radius of shadow, therefore one cannot distinguish them by the mass and the shadow size. But this does not mean they have the same intensity image of shadow. In fact, as we will also show, the conformal transformation has an impact on the BH shadow image.

The intensity image is affected not only by the conformal transformation of metric, but also by the accretion disk around the BH. This can be understood by analogy with the image of a candle flame formed by a lens. Here emissions from the disk play the role of the candle flame, and the central BH or ultra-compact star works as a very strong lens. The accretion disk can be thick or thin geometrically. In theoretical studies, optically thin and geometrically thick accretions are usually approximated by a spherical model, while optically thin emissions from geometrically thin accretions are often modeled as a thin disk.

In this paper, we will polish both models to study intensity images of the shadow of the conformal Schwarzschild BH. The organization is as follows. In Sec. \ref{subsect-conSch}, we specify the metric of the conformal Schwarzschild BH under study. In Sec. \ref{subsect-isco}, we calculate its radii of photon sphere, shadow and ISCO, and demonstrate that the conformal Schwarzschild BH has the same shadow size as the Schwarzschild BH if their masses are equal. In Sec. \ref{subsect-emiss}, we polish the emission models of spherical accretions in Ref. \cite{Narayan:2019imo} and thin disk accretions in Ref. \cite{Gralla:2019xty}. Shadow images of BHs surrounded by spherical accretions and thin disk accretions are simulated in Sec. \ref{subsect-sph} and Sec. \ref{subsect-disk} respectively. The results are compared against observations of the Event Horizon Telescope in Sec. \ref{subsect-obs}. We present our conclusions in Sec. \ref{sect-con}.

\section{The models}\label{sect-model}
\subsection{Nonsingular black holes in conformal gravity}\label{subsect-conSch}
It is generally expected that the theory of quantum gravity can get rid of classical singularities and quantum divergences in the Einstein theory. However, there is no consensus on how to get rid of them. By far the most promising candidate theory of quantum gravity is the superstring theory, in which the worldsheet conformal symmetry plays a crucial role. Unfortunately, to avoid conformal anomalies, the spacetime dimension should be $26$, $10$ or $2$ with $N=0$, $1$ or $2$ worldsheet supersymmetries. If alternatively the conformal symmetry is introduced to the 4-dimensional spacetime rather than the 2-dimensional worldsheet, is it possible to get rid of classical singularities and quantum divergences? Ref. \cite{Modesto:2016max} provided some evidence for a confirmative answer to this question. Concretely, Ref. \cite{Modesto:2016max} proposed a conformally invariant gravitational theory and obtained a class of BH solutions
\begin{equation}\label{metric-conSch}
ds^2=S(r)\left[-\left(1-\frac{\rh}{r}\right)dt^{2}+\left(1-\frac{\rh}{r}\right)^{-1}dr^{2}+r^2\dom\right]
\end{equation}
via a conformal transformation of the Schwarzschild metric. We will briefly refer to such a BH as a conformal Schwarzschild BH.

Here the scale factor can be taken as
\begin{equation}\label{S}
S(r)=1+\frac{L^4}{r^4}
\end{equation}
so that the spacetime is free of intrinsic singularity \cite{Modesto:2016max}. Especially, this spacetime is geodesically complete, and its Ricci scalar and Kretschmann scalar are regular everywhere \cite{Bambi:2016wdn}. In our notational convention, $L$ is a length scale prescribed by the gravitational theory \cite{Modesto:2016max,Bambi:2016wdn}, while $\rh$ is the Schwarzschild radius related to the BH mass $M$ by $\rh=2GM/c^2$. In the present paper, we will study three benchmark models with  $L=0$, $\rh$ and $1.4\rh$ in Eq. \eqref{S}, though there are many other choices for $S(r)$ and $L$ \cite{Modesto:2016max,Bambi:2017yoz}. Apparently, when $L=0$, the scale factor \eqref{S} reduces to $S=1$, and the metric \eqref{metric-conSch} goes back to the familiar Schwarzschild metric. In the limit $M=0$, the three models all go back to the flat spacetime. From the point of view of Refs. \cite{Modesto:2016max,Bambi:2016wdn,Bambi:2017yoz}, different choices of function $S(r)$ correspond to different choices of gauge before the conformal symmetry is broken. After symmetry breaking, they correspond to different vacua and can produce observable effects.

\subsection{Photon sphere, shadow and innermost stable circular obit}\label{subsect-isco}
For brevity and generality, let us write the metric of a static spherical BH in the general form
\begin{equation}\label{metric-sph}
ds^2=-A(r)dt^{2}+B(r)dr^{2}+D(r)\dom.
\end{equation}
In the following, general formulas will be written in terms of $A$, $B$ and $D$ and then applied to the specific case Eq. \eqref{metric-conSch}, which corresponds to
\begin{equation}\label{ABD}
A(r)=\left(1+\frac{L^4}{r^4}\right)\left(1-\frac{\rh}{r}\right),~~~~B(r)=\left(1+\frac{L^4}{r^4}\right)\left(1-\frac{\rh}{r}\right)^{-1},~~~~D(r)=r^2+\frac{L^4}{r^2}.
\end{equation}

For a spherical BH described by Eq. \eqref{metric-sph}, the radial coordinate of the photon sphere $\rph$ is the largest root of the equation \cite{Perlick:2021aok}
\begin{equation}\label{rph}
\frac{d}{dr}\left[\frac{D(r)}{A(r)}\right]=0.
\end{equation}
Observed at a distance of $\sqrt{D(\ro)}$, the angular radius of shadow $\ash$ satisfies
\begin{equation}\label{ash}
\sin^2\ash=\frac{D(\rph)A(\ro)}{A(\rph)D(\ro)},
\end{equation}
and the radius of shadow is \cite{Zhu:2021tgb}
\begin{equation}\label{Rsh}
\rsh=\sqrt{\frac{D(\rph)A(\ro)}{A(\rph)}}.
\end{equation}
Thanks to the cancellation of $S(r)$ in Eq. \eqref{rph}, the conformal Schwarzschild BH \eqref{metric-conSch} possesses a photon sphere of the same radial coordinate as the Schwarzschild BH, $\rph=3\rh/2$. As a result, seen by a distant observer, $A(\ro)\rightarrow1$, $D(\ro)\rightarrow\ro^2$, shadows of the conformal and the standard Schwarzschild BHs have the same angular radius, $\ash=3\sqrt{3}\rh/(2\ro)$, and the same radius, $\rsh=3\sqrt{3}\rh/2$.

\begin{figure}
\centering
\includegraphics[width=0.45\textwidth]{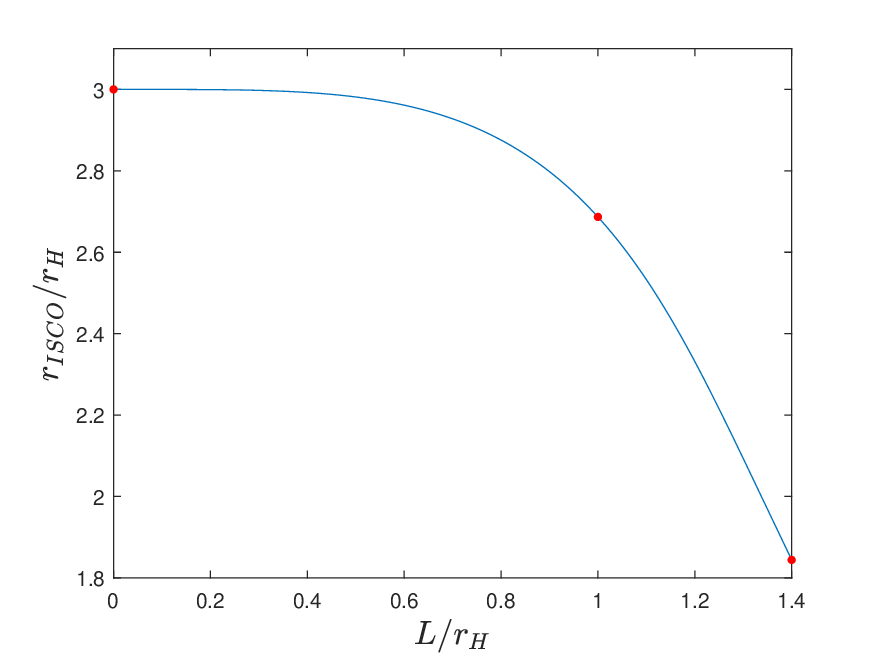}\\
\caption{The radius of ISCO calculated numerically from Eq. \eqref{isco}. Three benchmark cases $L/\rh=0$, $1$, $1.4$ are highlighted by red thick dots.}\label{fig-isco}
\end{figure}

In the spacetime \eqref{metric-sph}, the equations of motion of a massive particle on the equatorial plane $\theta=\pi/2$ can be integrated to give the conserved energy and angular momentum per unit mass,
\begin{equation}\label{el}
A(r)\frac{dt}{d\tau}=\varepsilon,~~~~D(r)\frac{d\phi}{d\tau}=\ell
\end{equation}
with $\tau$ being the proper time.\footnote{We follow Refs. \cite{Modesto:2016max,Bambi:2016wdn,Bambi:2017yoz} to consider conformally invariant gravitational theories minimally (but nonconformally) coupling to photons and matter in a certain frame. The true orbits of motion of photons and matter are determined by their minimal couplings to gravity in this frame. The scenario is different from the conformally invariant gravitational theories conformally (but nonminimally) coupling to photons and matter advocated in Refs. \cite{Mannheim:1991ez,Hobson:2022ahe,Mannheim:2021fql}.} Substituting them into the normalization condition of the 4-velocity, $g_{\mu\nu}u^{\mu}u^{\nu}=-1$, we get \cite{Song:2021ziq}
\begin{equation}\label{V}
A(r)B(r)\left(\frac{dr}{d\tau}\right)^2=\varepsilon^2-V(r),~~~~V(r)=A(r)\left[1+\frac{\ell^2}{D(r)}\right].
\end{equation}
One can check that it recovers the standard form when restricted to the Schwarzschild BH \cite{Carroll:2004st}. In general static spherically symmetric spacetimes, the ISCO can be determined from the effective potential by $V'(r)=V''(r)=0$ \cite{Bambi:2017yoz,Song:2021ziq}, that is $A''(A/D)'=A'(A/D)''$. Throughout this paper, the prime denotes differentiation with respect to $r$. Applying it to Eq. \eqref{metric-conSch}, we find $S(r)$ is not fully cancelled out, and the radius of ISCO is dictated by
\begin{equation}
\left(\frac{2S''}{r^3}+\frac{6S'}{r^4}\right)\cdot1+\left(-\frac{5S''}{r^3}-\frac{14S'}{r^4}+\frac{2S}{r^5}\right)\frac{\rh}{r}+\left(\frac{3S''}{r^3}+\frac{6S'}{r^4}-\frac{6S}{r^5}\right)\frac{\rh^2}{r^2}=0.
\end{equation}
If the scale factor $S(r)$ is not a constant, the root of this equation will deviate from that of the Schwarzschild BH. When it is specified as Eq. \eqref{S}, the above equation becomes
\begin{equation}\label{isco}
\frac{16L^4}{r^4}\cdot1+\left(2-\frac{42L^4}{r^4}\right)\frac{\rh}{r}+\left(-6+\frac{30L^4}{r^4}\right)\frac{\rh^2}{r^2}=0.
\end{equation}
We have numerically solved this equation and illustrated $\risco$ in Fig. \ref{fig-isco}. As highlighted by red thick dots, in three specific cases $L/\rh=0$, $1$, $1.4$, we have $\risco/\rh\approx3$, $2.69$, $1.84$ accordingly.\footnote{Inclusion of the term $\partial V'_{l_o}(r_o)/\partial l_o$ proposed in Ref. \cite{Song:2021ziq} would make a negligible contribution our numerical results.} Introducing a new variable $u=\rh/r$, we can rewrite Eq. \eqref{isco} as
\begin{equation}
\frac{L}{\rh}=\frac{1}{u}\left(\frac{6u^2-2u}{16-42u+30u^2}\right)^{1/4},
\end{equation}
which attains its maximum $L/\rh=3/2$ at $u=2/3$. It means the ISCO does not exist unless $L/\rh\leq3/2$. As partly shown in Fig. \ref{fig-isco}, $\risco/\rh$ is a monotonic function of $L/\rh$, hence we can also say that $\risco$ attains its minimum $\risco=3\rh/2$ at $L/\rh=3/2$.
\begin{figure}
\centering
\includegraphics[width=0.45\textwidth]{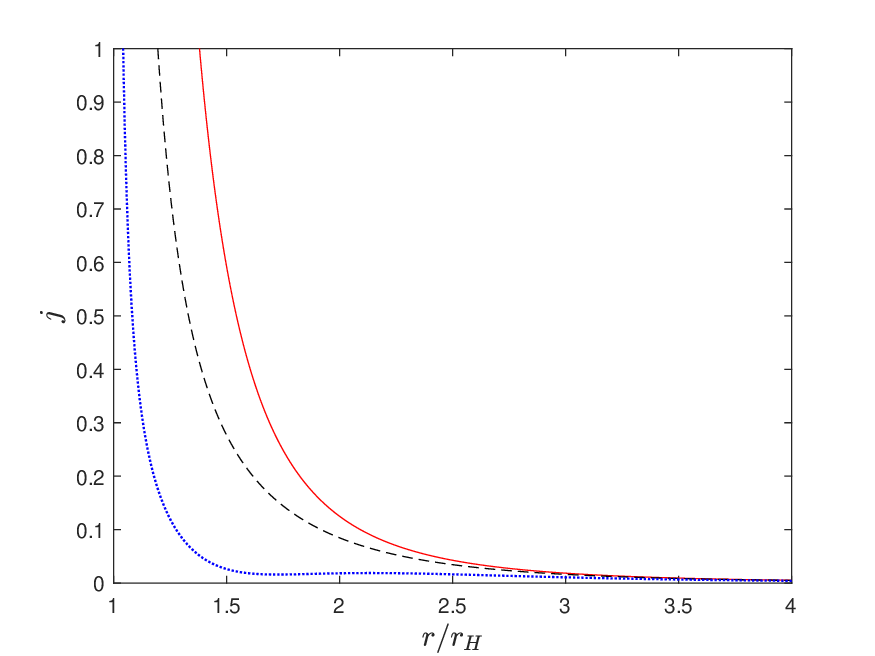}\\
\caption{The emission coefficient per unit solid angle following Eq. \eqref{j} without an inner cutoff. Three benchmark models $L/\rh=0$, $1$, $1.4$ are illustrated by red solid, black dashed, blue dotted curves respectively.}\label{fig-jem}
\end{figure}

\subsection{From emission coefficient to observed intensity}\label{subsect-emiss}
The equations of motion of photons (massless particles) on the equatorial plane $\theta=\pi/2$ take the same form as Eqs. \eqref{el}, \eqref{V}, except for that the effective potential is now
\begin{equation}\label{Vph}
V(r)=\frac{\ell^2A(r)}{D(r)}
\end{equation}
and the conserved quantities $\varepsilon$, $\ell$ are the energy and angular momentum of photons. Defining an impact parameter $b=\ell/\varepsilon$, we can combine them as an orbital equation
\begin{equation}\label{orbitr}
\frac{dr}{d\phi}=\pm\sqrt{\frac{D(r)}{B(r)}\left[\frac{D(r)}{b^2A(r)}-1\right]}
\end{equation}
with the upper (lower) sign for photons approaching (leaving) the BH clockwisely. In virtue of the spherical symmetry, we will pay attention to clockwise light rays and trace the light rays backwards from the observer. Aiming at the photon sphere, the photon has a critical value of impact parameter \cite{Zhu:2021tgb}
\begin{equation}\label{bph}
\bph=\sqrt{\frac{D(\rph)}{A(\rph)}}.
\end{equation}
Thanks to the cancellation of $S(r)$ in the ratio $D/A$, the conformal Schwarzschild BH \eqref{metric-conSch} has the same critical impact parameter as the Schwarzschild BH, $\bph=3\sqrt{3}\rh/2$.


In the literature, various toy models of emission coefficient have been adopted for studying BH images. Among them the most popular ones are those designed in Ref. \cite{Zeng:2020vsj}, and similar models can be found in Refs. \cite{Peng:2020wun,Li:2021ypw}. We will not follow them here. Physically a more reasonable model seems to be the one proposed in Ref. \cite{Narayan:2019imo}, which assumes the total luminosity emitted between radius $r$ and infinity is proportional to the total potential energy drop from infinity to $r$. Basing on this assumption, we find the emission coefficient per unit solid angle
\begin{equation}\label{j}
j(r)=\int_0^\infty j(\nu_{\mathrm{e}},r)d\nu_{\mathrm{e}}=\left\{\begin{array}{ll}
0, & \hbox{$r\leq\rin$;} \\
\frac{A'(r)}{32\pi^2A(r)D(r)\sqrt{A(r)B(r)}}, & \hbox{$r>\rin$}
\end{array}\right.
\end{equation}
($\mathrm{erg}~\mathrm{cm}^{-3}~\mathrm{s}^{-1}~\mathrm{ster}^{-1}$) for a radiating gas at rest in the spherical spacetime \eqref{metric-sph}. When deriving this expression, we have taken the energy drop to be $\sqrt{A(\infty)}-\sqrt{A(r)}$, which can be inferred from Eqs. \eqref{V} by setting the kinetic energy to zero. In Fig. \ref{fig-jem}, we have illustrated the emission coefficient for three benchmark models mentioned in Sec. \ref{subsect-conSch}. In the next section, we will simulate images of BHs in two accretion models, i.e., the spherical model and the thin disk model, each with three different values of inner boundary radius $\rin=\rh$, $\rph$, $\risco$. For thin disk accretions, Eq. \eqref{j} should be understood as the emission coefficient on the disk, but no emission is expected elsewhere.

In this paper, we will consider emissions from a free-falling gas. If $A(r)$ tends to 1 at the spatial infinity, it can be demonstrated that the redshift factor for a distant observer is
\begin{equation}\label{g}
g=\frac{\nu_{\mathrm{obs}}}{\nu_{\mathrm{e}}}=\frac{A(r)}{1\pm\sqrt{\left[1-A(r)\right]\left[1-\frac{b^2A(r)}{D(r)}\right]}}
\end{equation}
with the upper (lower) sign for photons leaving (approaching) the BH \cite{Bambi:2013nla}, which can be understood as the Doppler deboosting (boosting) effect. Then in the observed BH image, the bolometric intensity at the point of impact parameter $b$ is
\begin{equation}\label{Iint}
I(b)=\int_{\mathrm{ray}}g^4j(r)\dl.
\end{equation}
For spherical accretions, the infinitesimal proper length is simply
\begin{equation}\label{dlr}
\dl=\sqrt{B(r)+D(r)\left(\frac{d\phi}{dr}\right)^2}dr,
\end{equation}
which can be inserted into Eq. \eqref{Iint} to perform the integration.

For a BH surrounded by a thin accretion disk, its image has a dependence on the inclination between the line of sight and the disk axis. In our study, we will consider the face-on case, in which the line of sight is perpendicular to the plane of the accretion disk. We take the disk to lie on the prime vertical plane $\phi=\pi/2$ or $3\pi/2$, while the observer is located in the direction $\phi=0$ on the equatorial plane $\theta=\pi/2$. This configuration is the same as the one in Sec. IIB of Ref. \cite{Gralla:2019xty}, but in a different spherical coordinate system. By virtue of the circular symmetry, it is enough to study photons moving on the equatorial plane. All of them are emitted from the intersection of the equatorial plane and the disk. Therefore, neglecting the thickness of disk, one should insert a factor $\sum_m\delta(\phi+\pi/2-m\pi)$ into the integrand of Eq. \eqref{Iint} and set $\theta=\pi/2$. For this reason, it is convenient to rewrite the infinitesimal proper length as
\begin{equation}\label{dlphi}
\dl=\sqrt{B(r)\left(\frac{dr}{d\phi}\right)^2+D(r)}d\phi,
\end{equation}
which enables us to work out the integration \eqref{Iint}, arriving at a sum
\begin{equation}\label{Isum}
I(b)=\sum_m \left.g^4j(r)\frac{D(r)}{b\sqrt{A(r)}}\right|_{r=r_m(b)}.
\end{equation}
In this expression, $r_m$ is evaluated at the $m$-th intersection of the backward trajectories of photon and the disk \cite{Gralla:2019xty}. Note here is a new factor $D/(b\sqrt{A})$ as compared with Eq. (12) in Ref. \cite{Gralla:2019xty}.

\section{The images}\label{sect-image}
\begin{figure}
\centering
\includegraphics[width=0.45\textwidth]{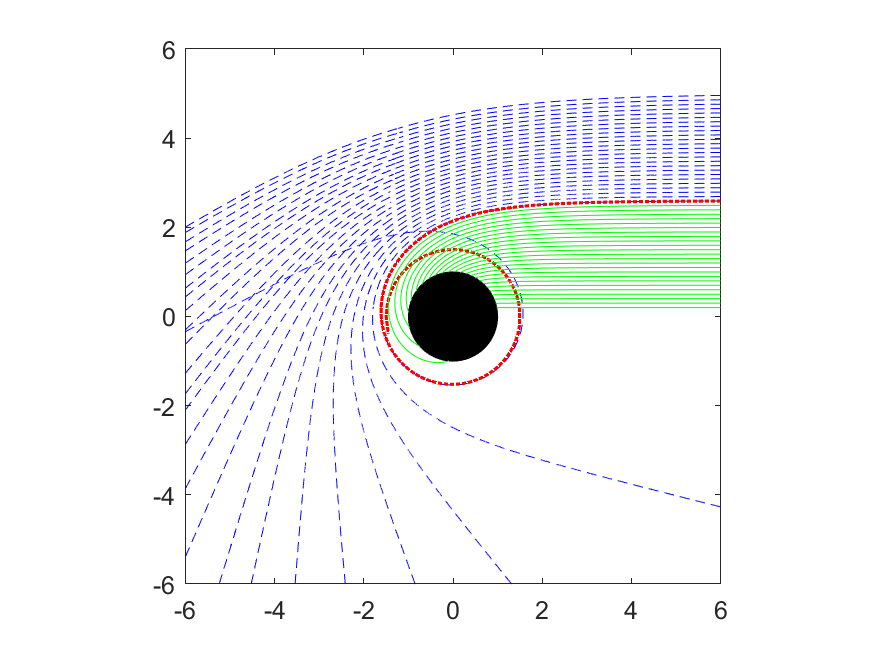}\includegraphics[width=0.45\textwidth]{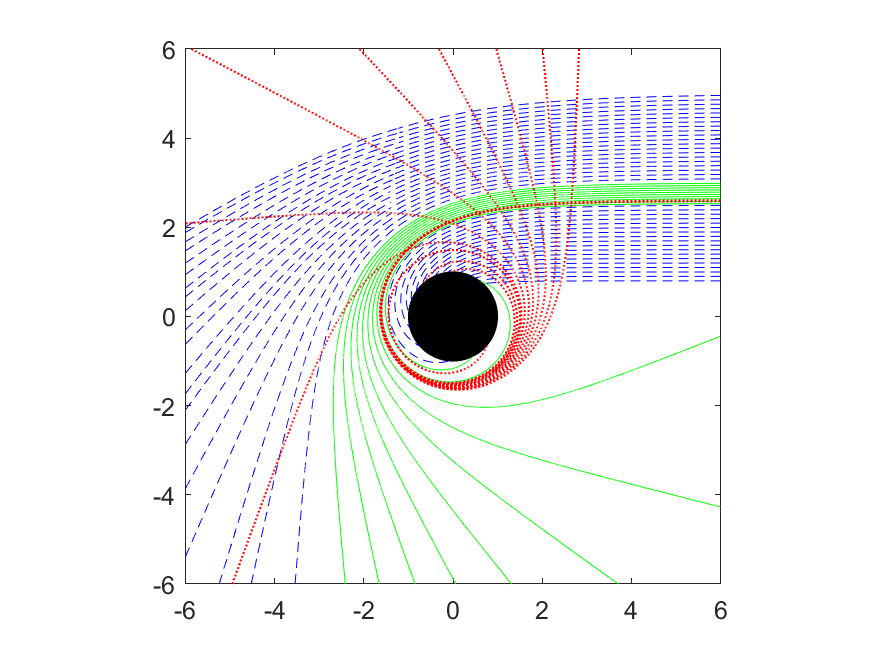}\\
\caption{Trajectories of the light rays simulated according to Eq. \eqref{orbitr} and their classifications in Sec. \ref{subsect-sph} and Sec. \ref{subsect-disk}. All axes are rescaled by $\rh$. The BH is marked by black disks. Left panel: green solid curves, red dotted curves and blue dashed curves depict trajectories with $b<\bph$, $b=\bph$ and $b>\rph$, respectively. Right panel: blue dashed curves, green solid curves and red dotted curves depict direct, lensed and photon-ring trajectories, respectively.}\label{fig-traj}
\end{figure}

Having established our models, we are now ready to draw intensity images of the BH shadow cast. The images are formed by photons arriving at a distant observer after bypassing the BH. Outside the event horizon, we work with the variable $u=\rh/r$ which ranges in a finite interval naturally, then the photon trajectories can be obtained numerically from the orbital equation \eqref{orbitr}, see Fig. \ref{fig-traj}. As elaborated above, we are interested in images of BHs in two accretion models: the spherical model and the thin disk model. For both accretion models, we will study three benchmark cases $L/\rh=0$, $1$, $1.4$, and in every case three different values of inner radius $\rin=\rh$, $\rph$, $\risco$. Therefore, the intensity images can be arranged in two sudoku figures, see Figs. \ref{fig-IS} and \ref{fig-ID}. The intensities of BHs surrounded by spherical accretions will be explored in Sec. \ref{subsect-sph}. They are simulated by performing the integration \eqref{Iint} in the backward ray-tracing method \cite{Luminet:1979nyg}. Intensities of BHs surrounded by thin disk accretions will be investigated in Sec. \ref{subsect-disk}. They are obtained by evaluating the sum \eqref{Isum} directly, where terms of $m\geq4$ are neglected. Throughout this section, we assume the emission coefficient is always given by Eq. \eqref{j}, and the accreting gas is free-falling, inducing a redshift factor \eqref{g}. Remarkably, there is a $g^4$ factor in both Eq. \eqref{Iint} and Eq. \eqref{Isum}, which will play an crucial role in the intensity profiles and images of BHs. In Sec. \ref{subsect-obs}, we will compare predictions of our benchmark models against observational data.

\subsection{Spherical accretions}\label{subsect-sph}
Narayan et. al. studied shadow images of the Schwarzschild BH surrounded by spherical accretions in Ref. \cite{Narayan:2019imo}. In Eq. \eqref{j}, we have suggested a relativistic extension of the emission coefficient in their model. In the current subsection, we adapt the procedure of Ref. \cite{Narayan:2019imo} to this polished model of spherical accretion. Details of the procedure can be found in Ref. \cite{Zeng:2020vsj}. We solve the photon trajectories from Eq.\eqref{orbitr} and classify them into three types: the unstable circular trajectories with $b=\bph$ escaping to infinity (deboosted) under perturbations, the trajectories with $b<\bph$ penetrating the photon sphere from inside to infinity (deboosted), and the symmetric trajectories with $b>\bph$ always outside the photon sphere. They are plotted in the left panel of Fig. \ref{fig-traj} by red dotted curves, green solid curves and blue dashed curves, respectively. Note that along the third type of trajectories, photons first approach the BH (boosted) and then leave it (deboosted), so there is no confusion about the sign in the redshift factor \eqref{g} for each type of trajectories. Fixing $L/\rh$, the trajectory is in one-to-one correspondence with the value of $b$. Therefore, given the values of $L/\rh$ and $\rin$, we can insert Eqs. \eqref{ABD}, \eqref{orbitr}, \eqref{j}, \eqref{g} and \eqref{dlr} into Eq. \eqref{Iint} to compute the intensity for each value of $b$ numerically, with the integration performed in terms of the variable $u=\rh/r$ using the backward ray-tracing method.

\begin{figure}
\centering
\hspace{0.1\textwidth}\includegraphics[width=0.36\textwidth]{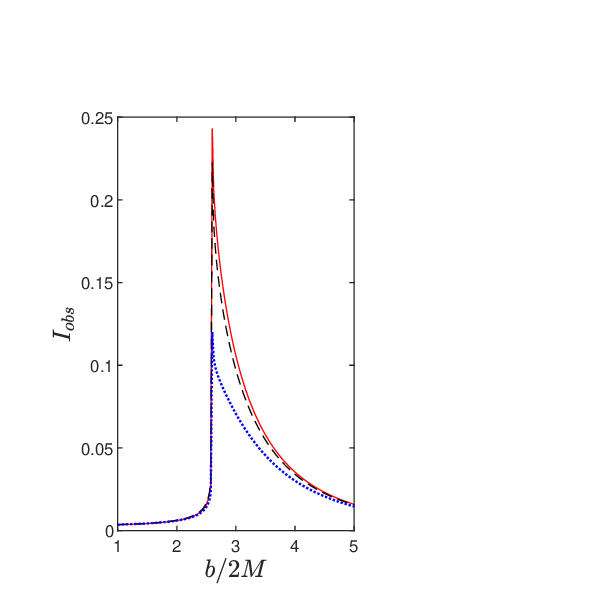}\hspace{-0.13\textwidth}\includegraphics[width=0.36\textwidth]{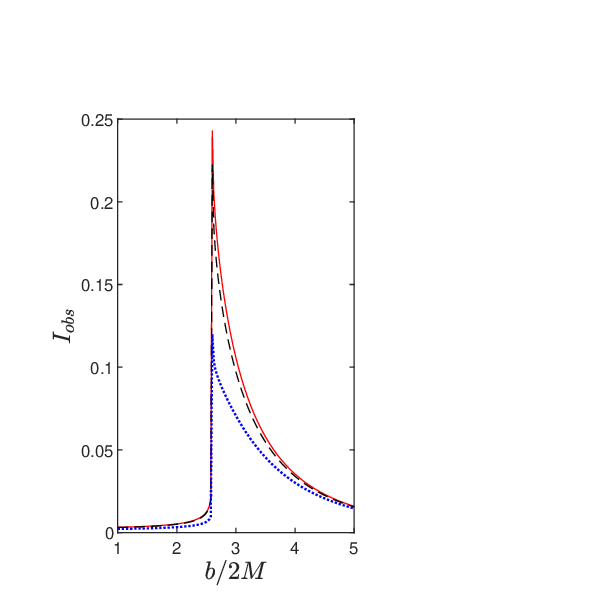}\hspace{-0.13\textwidth}\includegraphics[width=0.36\textwidth]{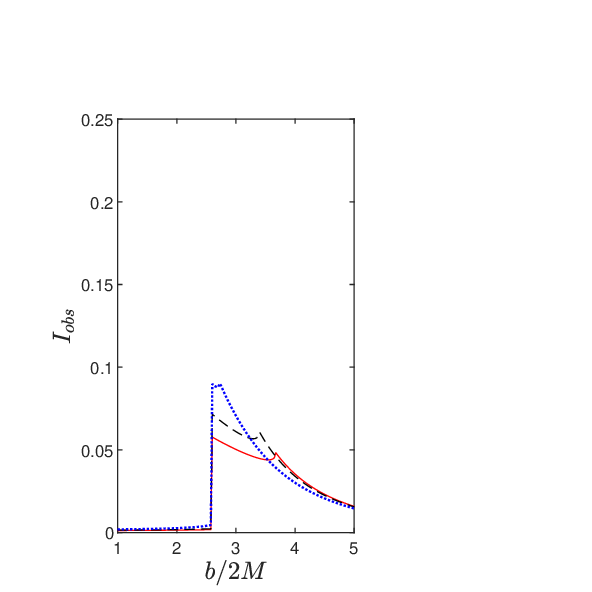}\\
\caption{Intensity profile $I(b)$ as a function of the impact parameter $b$ in the spherical accretion model. The radius of inner boundary is set to $\rin=\rh$, $\rph$, $\risco$ from left to right panels. In each panel, three benchmark models $L/\rh=0$, $1$, $1.4$ are illustrated by red solid, black dashed, blue dotted curves.}\label{fig-PS}
\end{figure}
\begin{figure}
\centering
\includegraphics[width=0.33\textwidth]{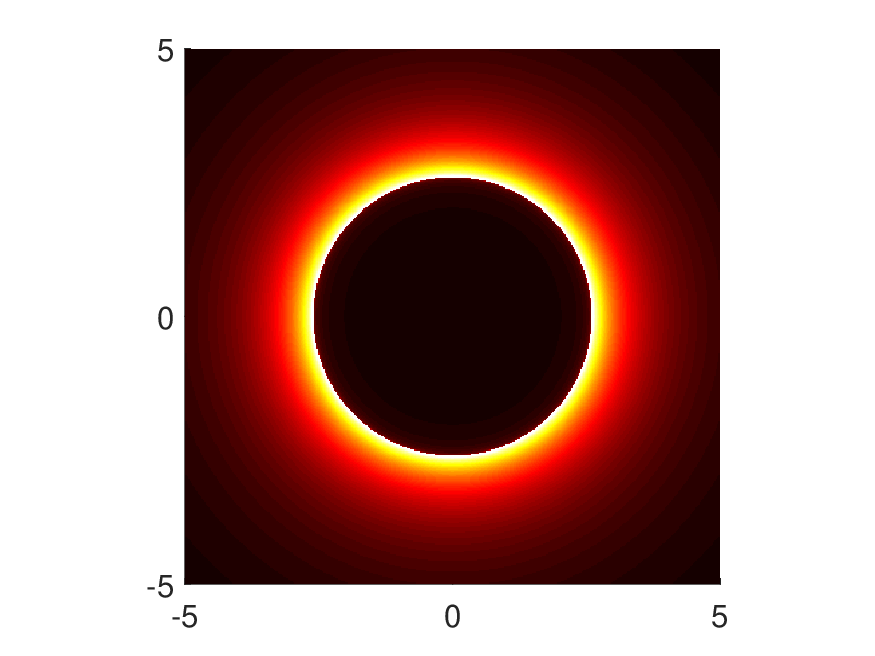}\hspace{-0.06\textwidth}\includegraphics[width=0.33\textwidth]{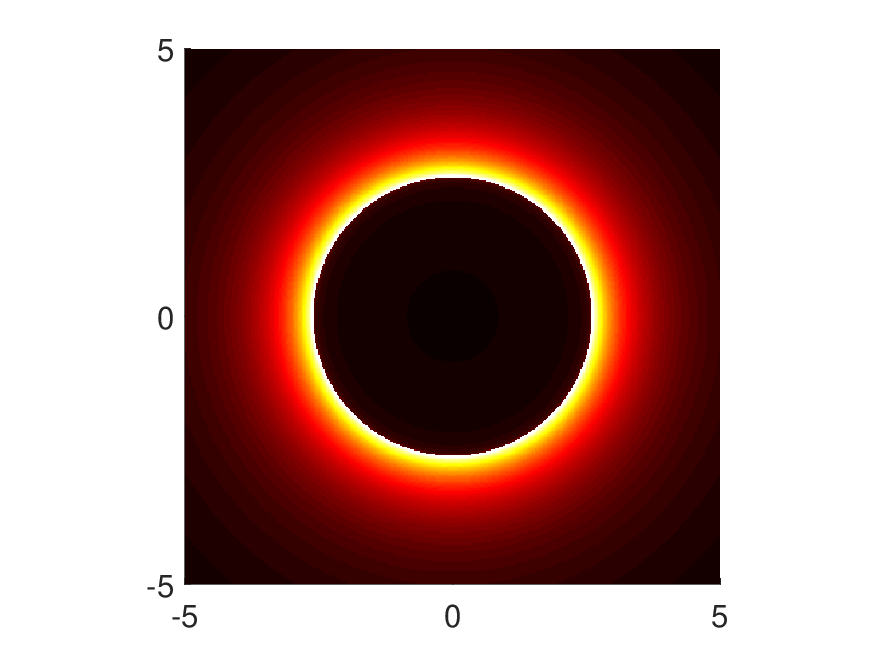}\hspace{-0.05\textwidth}\includegraphics[width=0.33\textwidth]{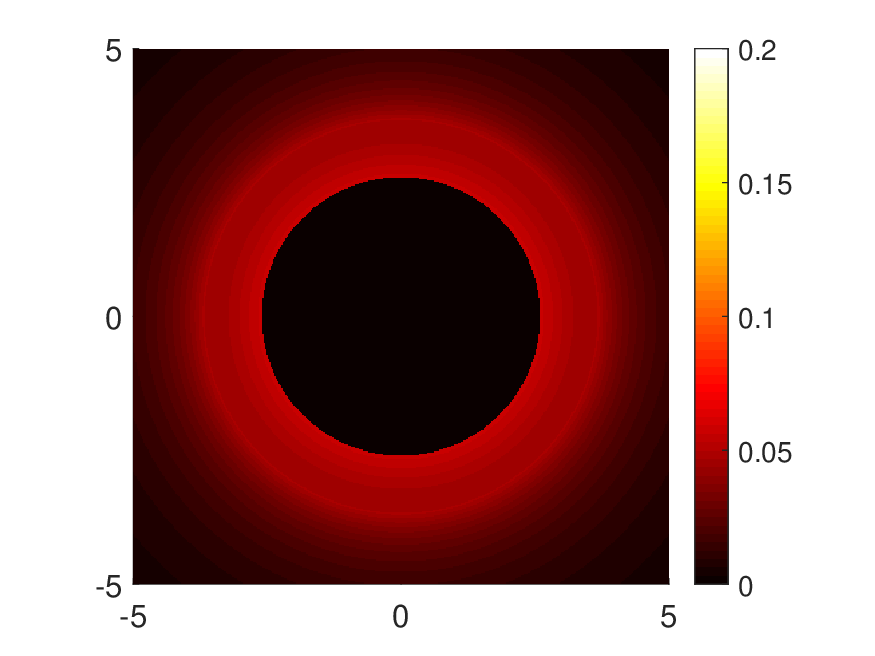}\\
\includegraphics[width=0.33\textwidth]{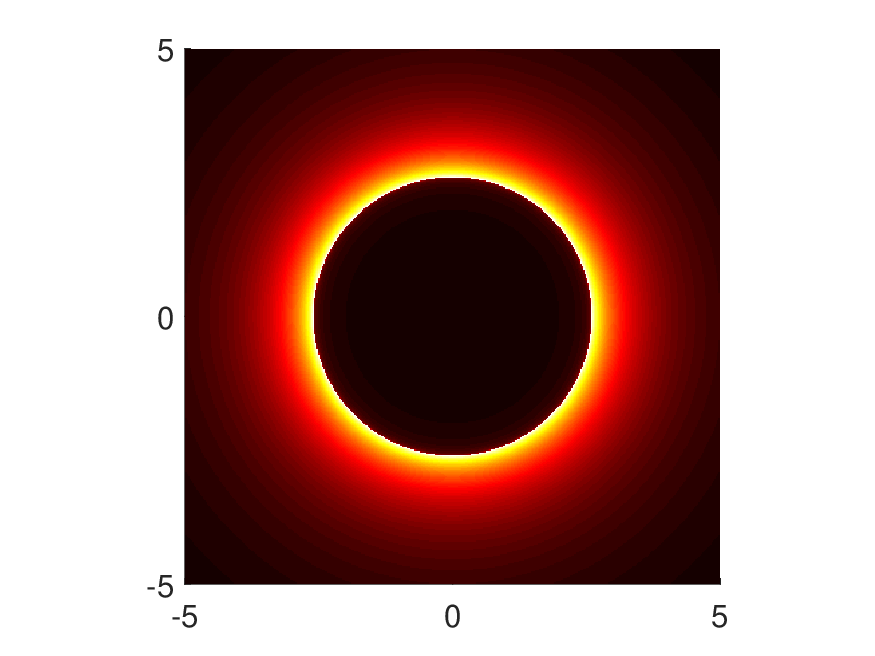}\hspace{-0.06\textwidth}\includegraphics[width=0.33\textwidth]{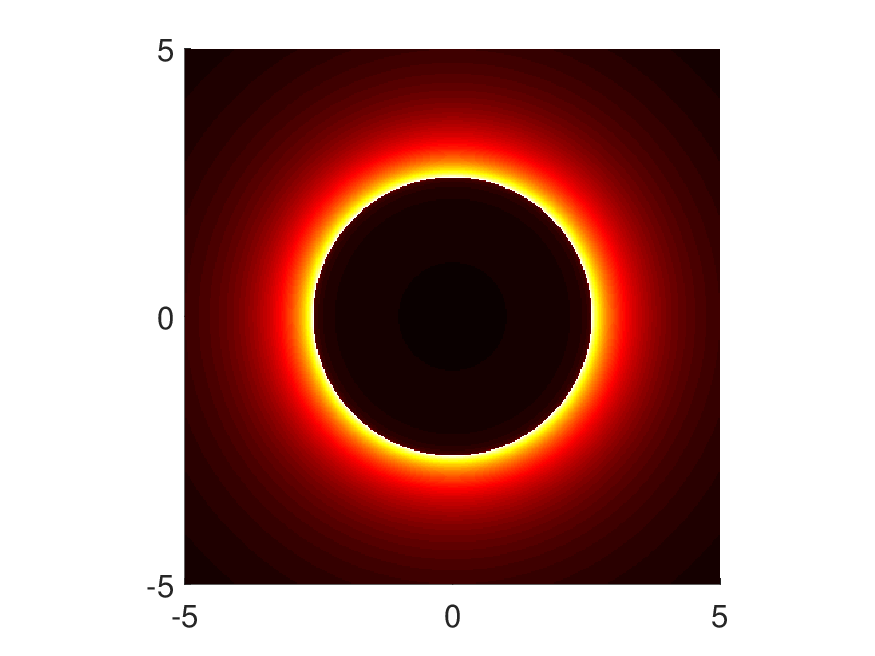}\hspace{-0.05\textwidth}\includegraphics[width=0.33\textwidth]{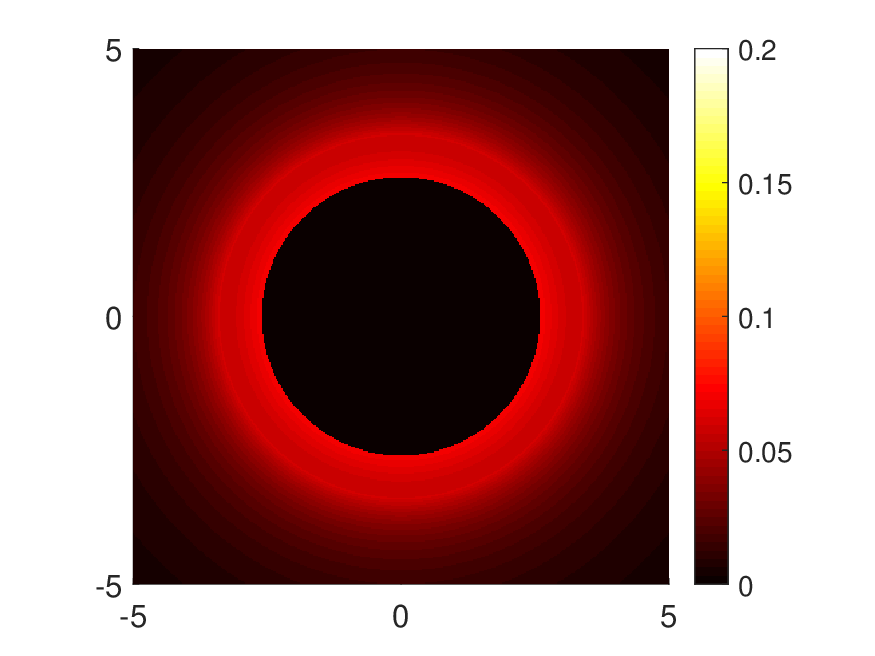}\\
\includegraphics[width=0.33\textwidth]{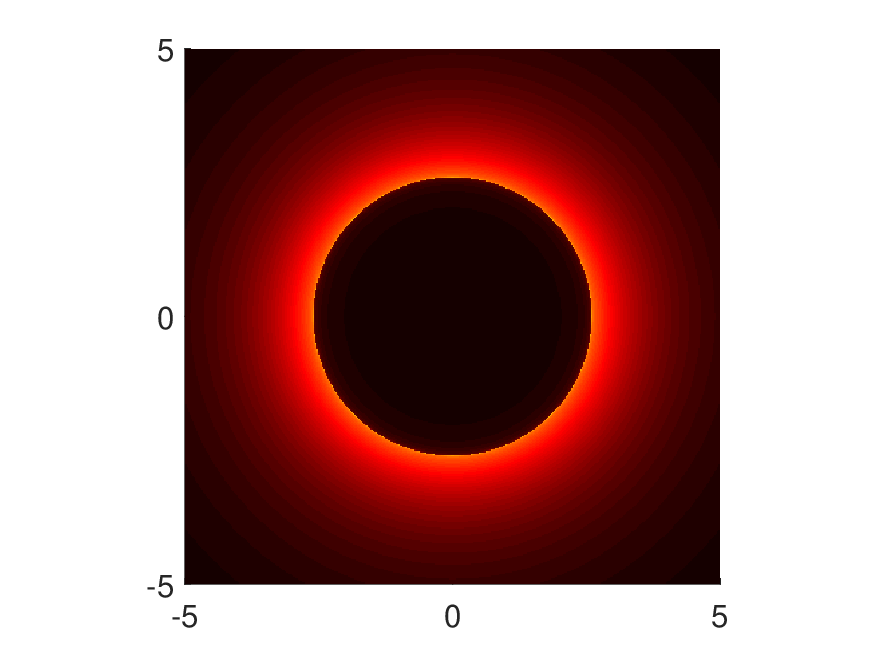}\hspace{-0.06\textwidth}\includegraphics[width=0.33\textwidth]{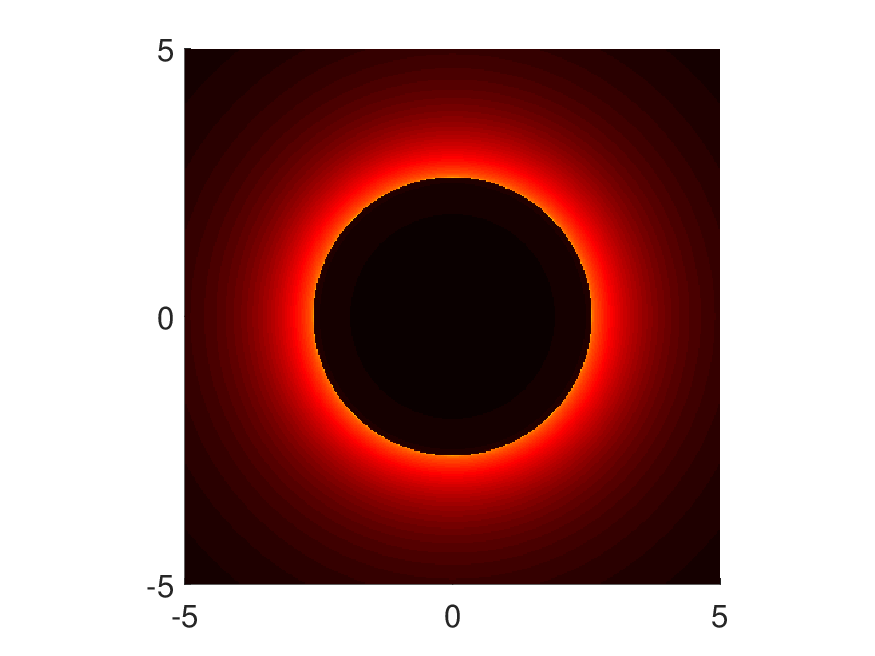}\hspace{-0.05\textwidth}\includegraphics[width=0.33\textwidth]{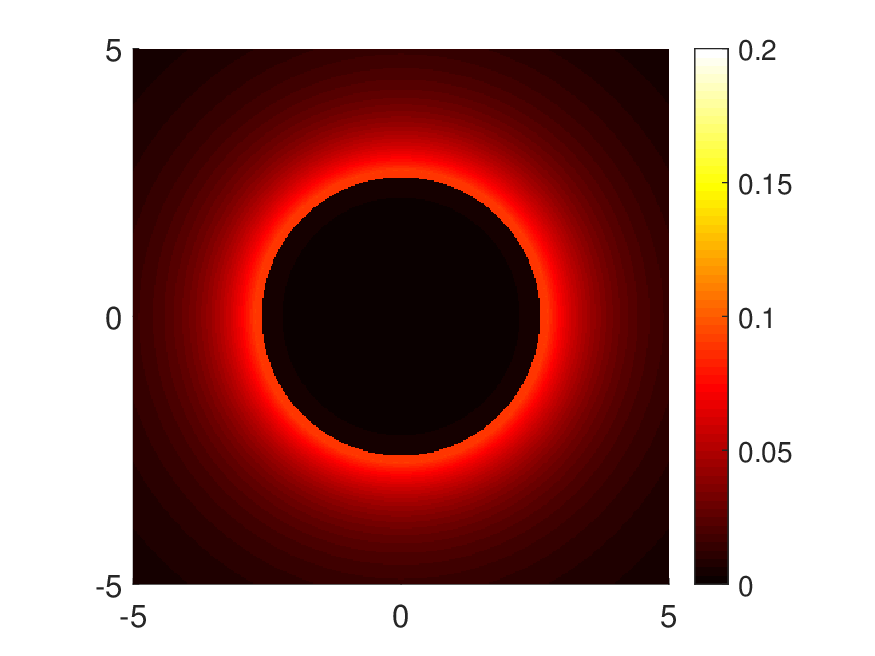}\\
\caption{Intensity images of the BH shadow cast in the spherical accretion model. We set the inner radius to $\rin=\rh$, $\rph$, $\risco$ from left to right panels, and the length parameter $L/\rh=0$, $1$, $1.4$ from top to bottom panels. All axes are rescaled by $\rh$.}\label{fig-IS}
\end{figure}

Our results are depicted in Figs. \ref{fig-PS} and \ref{fig-IS}. In the Schwarzschild limit $L/\rh=0$, the intensity profiles are shown by red solid curves in Fig. \ref{fig-PS}, and the intensity images by top panels in Fig. \ref{fig-IS}, both with $\rin/M=2$, $3$, $6$ from left to right. The profiles and images are consistent with Figs. 4 and 5 in Ref. \cite{Narayan:2019imo}. Particularly, because photons with $b\leq\bph$ are always deboosted but never boosted, the boundary of shadow is always located at the photon sphere irrespective of inner boundary radius of the accretion/emission.

When the length parameter is equal to the radial coordinate of the event horizon, $L/\rh=1$, the intensity profiles are demonstrated by black dashed curves in Fig. \ref{fig-PS}, and the intensity images by the middle row of Fig. \ref{fig-IS}, with $\rin/M=2$, $3$, $5.37$ from left to right. In the third case $L/\rh=1.4$, the intensity profiles are illustrated by blue dotted curves in Fig. \ref{fig-PS}, and the intensity images by bottom panels in Fig. \ref{fig-IS}, where $\rin/M=2$, $3$, $3.69$ from left to right. In both cases, the boundary of shadow is located at the photon sphere irrespective of the value of $\rin$.

From Figs. \ref{fig-PS} and \ref{fig-IS}, we can examine the influence of the conformal scale factor $S(r)$, or specifically, the length parameter $L$ on the intensity distribution of BH shadow. When $\rin=\rh$ or $\rin=\rph$, the contrast between the two sides of the photon sphere decreases as the ratio $L/\rh$ increases. On the contrary, when $\rin=\risco$, the contrast is enhanced by the increment of $L/\rh$.

In the right panel of Fig. \ref{fig-PS}, there is a prominent plateau in each curve. This is a consequence of the cutoff of emission coefficient \eqref{j} at $\rin=\risco$. As we have explained, photons with $b\leq\bph$ experience a Doppler deboosting only. However, as can be seen in Fig. \ref{fig-traj}, some photons with $b>\bph$ first approach the BH (boosted) and then get away (deboosted). What is more, the deboosting is stronger when photons are closer to the BH. Therefore, photons with $b\leq\bph$ make a negligible contribution to the intensity, while the contributions from photons with $b>\bph$ are more important. Furtherly, the plateaus (rather than peaks) in Fig. \ref{fig-PS} are caused by the lack of emissions in the region $\rph<r<\risco$. Recall that the impact parameter of a photon is related to the pericenter radius of its trajectory by Eq. \eqref{orbitr} with $dr/d\phi=0$, i.e., $b=\sqrt{D(\rmin)/A(\rmin)}$. In agreement with the plateau edges in Fig. \ref{fig-PS}, photons with pericenter radius $\rmin=\risco$ has the impact parameter $b/\rh=3.7$, $3.4$, $2.7$ in three specific cases $L/\rh=0$, $1$, $1.4$, respectively.

We close this subsection with some remarks on the emission coefficient Eq. \eqref{j}. Compared with its nonrelativistic form in Ref. \cite{Narayan:2019imo}, such a relativistic version of emission coefficient does not modify the observed intensity of Schwarzschild BH significantly, but it can be applied to a broader class of spacetimes such as the conformal Schwarzschild BH. Remember that Eq. \eqref{j} is based on the conversion of the gravitational potential energy into the emission energy, therefore this expression of emission coefficient is nonnegative if and only if $A'(r)\geq0$. If $A(r)$ is given by Eq. \eqref{ABD}, then it is not hard to see
\begin{equation}\label{dA}
\frac{\rh^3r^2}{L^4}A'(r)=5u^4-4u^3+\frac{\rh^4}{L^4}\geq\frac{\rh^4}{L^4}-\left(\frac{3}{5}\right)^3
\end{equation}
in the interval $0<u\leq1$. To guarantee $A'(r)\geq0$, we should restrict to the parameter region $L/\rh\leq(5/3)^{3/4}\approx1.47$.

\subsection{Thin disk accretions}\label{subsect-disk}
In Ref. \cite{Gralla:2019xty}, Gralla et. al. put forward a thin-disk accretion model to investigate shadow images of the Schwarzschild BH. They discovered that, in addition to a photon ring located at the photon sphere, there is a lensing ring somewhat outside this sphere. Assuming the accretion disk emits isotropically in the rest frame of static worldlines, they demonstrated that the direct emission dominates over the lensing-ring emission, and the flux from photon ring is negligible, so the size of the dark central area is very much dependent on the emission model (e.g., the inner radius of the disk).

Our thin disk model of accretion explained in Sec. \ref{subsect-emiss} differs from theirs in three aspects. First of all, the accreting gas is free-falling rather than static, thus the redshift factor \eqref{g} here is more complicated than Eq. (10) in Ref. \cite{Gralla:2019xty}. In particular, it depends on the moving direction of photon. Secondly, the sum form of the observed intensity Eq. \eqref{Isum} is derived rigorously from its integral form \eqref{Iint}, and it has a new factor $D/(b\sqrt{A})$ as compared with Eq. (12) in Ref. \cite{Gralla:2019xty}. Thirdly, the emission coefficient \eqref{j} is not the same as the emitted intensities in Fig 5 of Ref. \cite{Gralla:2019xty}.

For a given value of $b$, we can integrate Eq. \eqref{orbitr} to compute $r_1(b)$ at $\phi=\pi/2$, $r_2(b)$ at $\phi=3\pi/2$ and $r_3(b)$ at $\phi=5\pi/2$ if they exist. Then the observed intensity $I(b)$ is obtained by substituting $r_1(b)$, $r_2(b)$, $r_3(b)$ into Eq. \eqref{Isum}. Following Ref. \cite{Gralla:2019xty}, we classify the photon trajectories into three types: the direct trajectories that intersect the plane of disk only once outside the horizon, the lensed trajectories that cross the plane of disk twice, and the photon-ring trajectories crossing the disk plane three or more times. They are plotted in the right panel of Fig. \ref{fig-traj} by blue dashed curves, green solid curves and red dotted curves, respectively. As clarified in Ref. \cite{Gralla:2019xty}, the direct emission contributes to the $m=1$ term in Eq. \eqref{Isum} only, whereas the lensing-ring emission contributes to both $m=1,2$ terms but not to others. The photon-ring emission contributes to $m=1,2,3$ and more terms. For each $m$, Eq. \eqref{orbitr} has a solution $r_m(b)$ if and only if $b$ is in the interval $(b_m^{-},b_m^{+})$. As $m$ increases, the interval becomes narrower and narrower \cite{Gralla:2019xty,Bisnovatyi-Kogan:2022ujt}. Consequently, contributions from terms of $m\geq4$ are negligible in practice.

\begin{figure}
\centering
\hspace{0.13\textwidth}\includegraphics[width=0.36\textwidth]{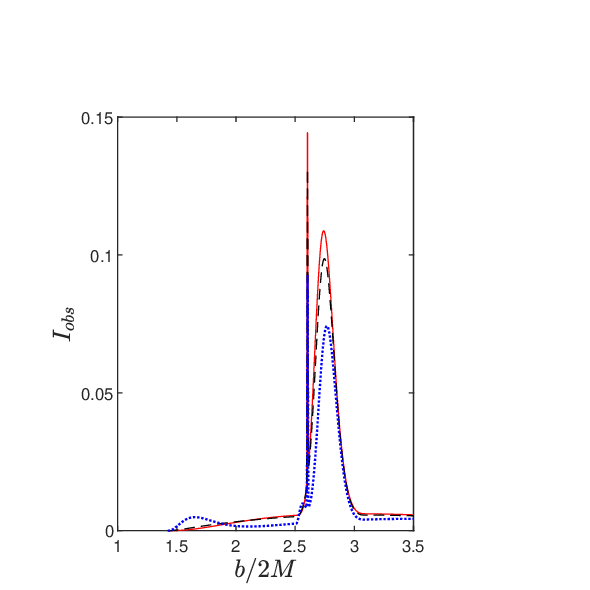}\hspace{-0.1\textwidth}\includegraphics[width=0.36\textwidth]{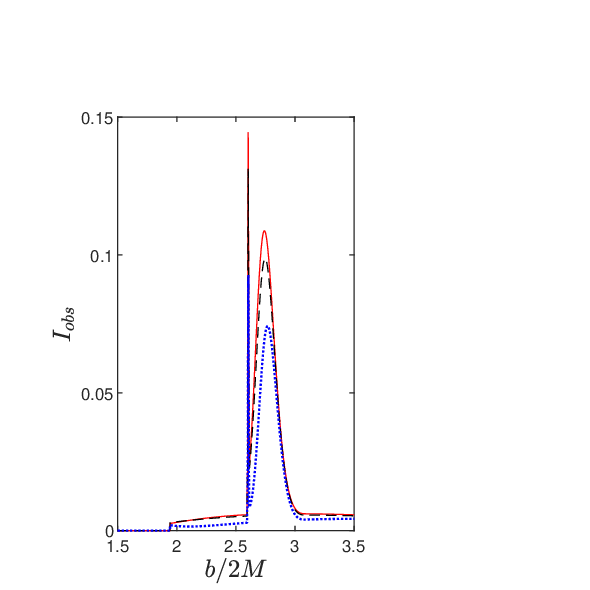}\hspace{-0.13\textwidth}\includegraphics[width=0.36\textwidth]{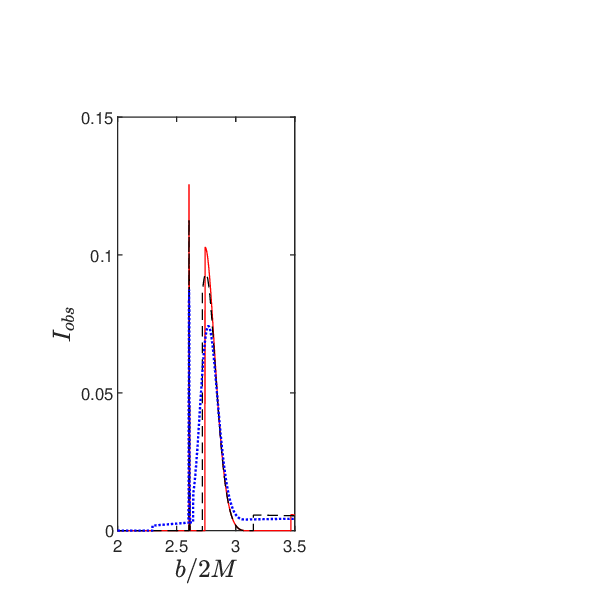}\\
\caption{Intensity profile $I(b)$ as a function of the impact parameter $b$ in the thin-disk accretion model. Same notations as in Fig. \ref{fig-PS}.}\label{fig-PD}
\end{figure}
\begin{figure}
\centering
\includegraphics[width=0.33\textwidth]{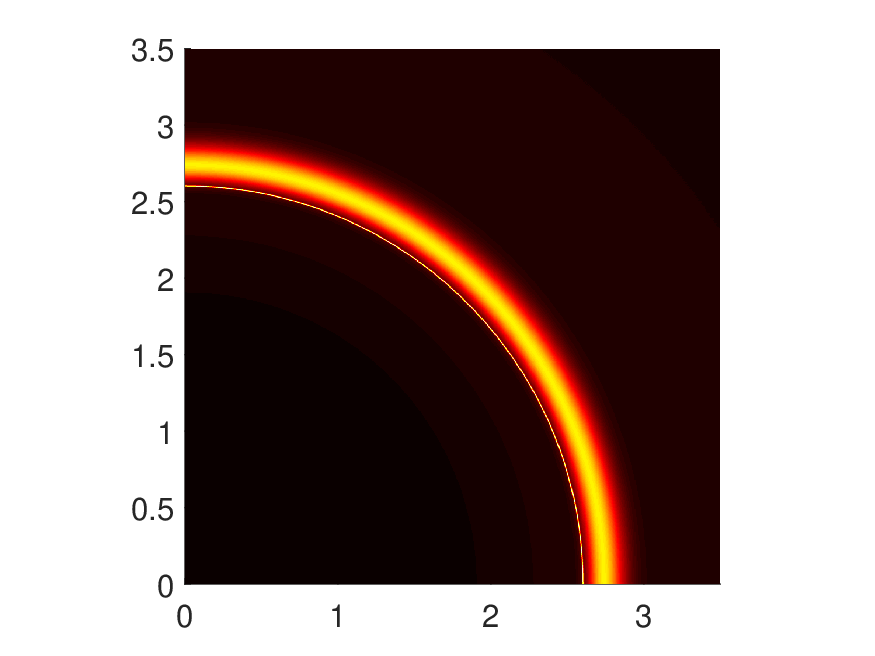}\hspace{-0.06\textwidth}\includegraphics[width=0.33\textwidth]{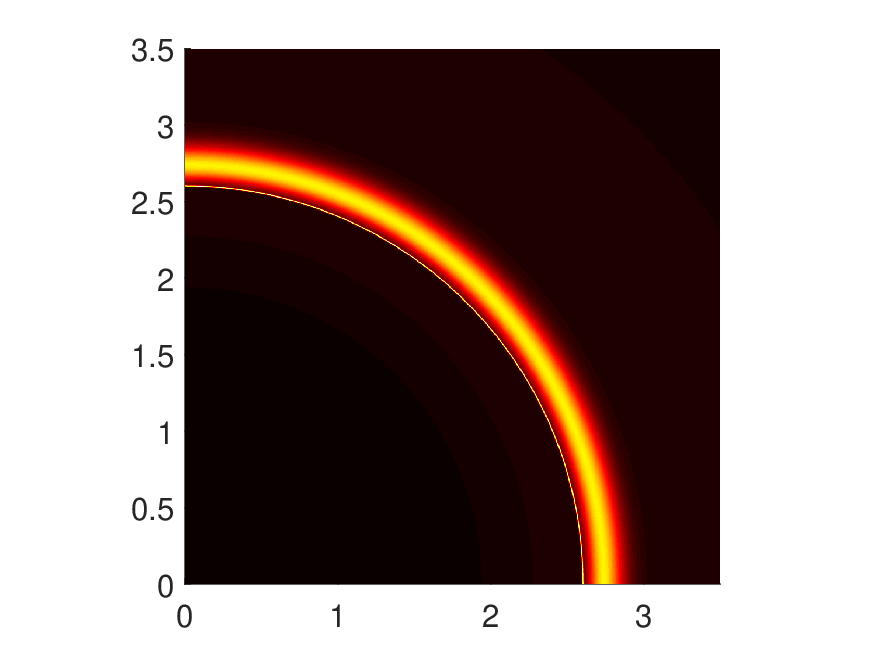}\hspace{-0.05\textwidth}\includegraphics[width=0.33\textwidth]{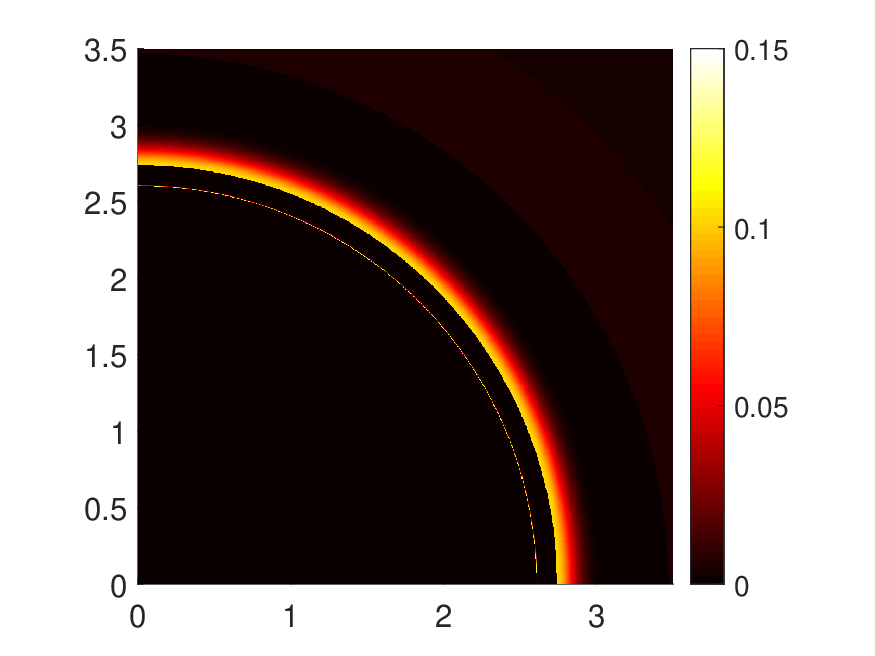}\\
\includegraphics[width=0.33\textwidth]{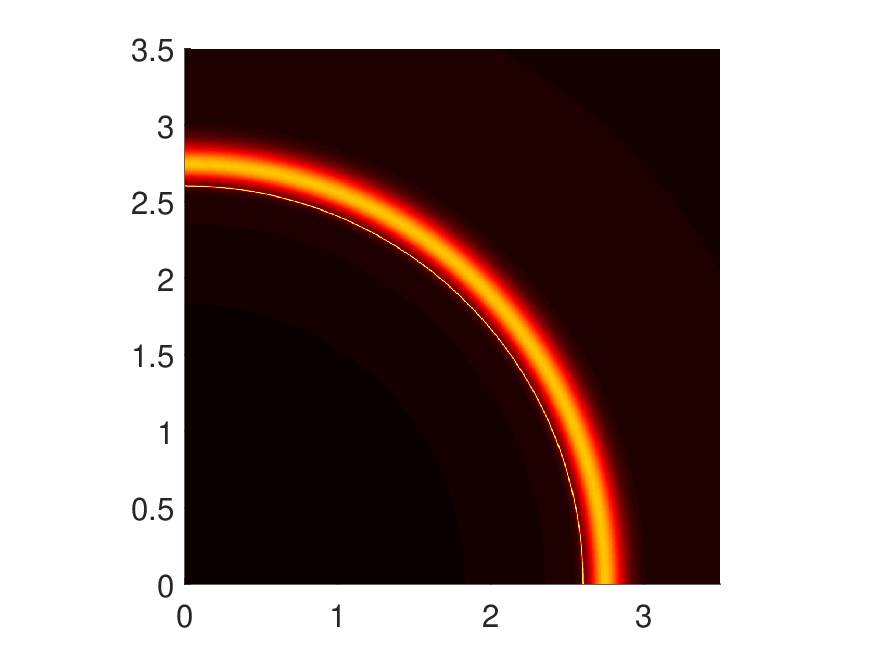}\hspace{-0.06\textwidth}\includegraphics[width=0.33\textwidth]{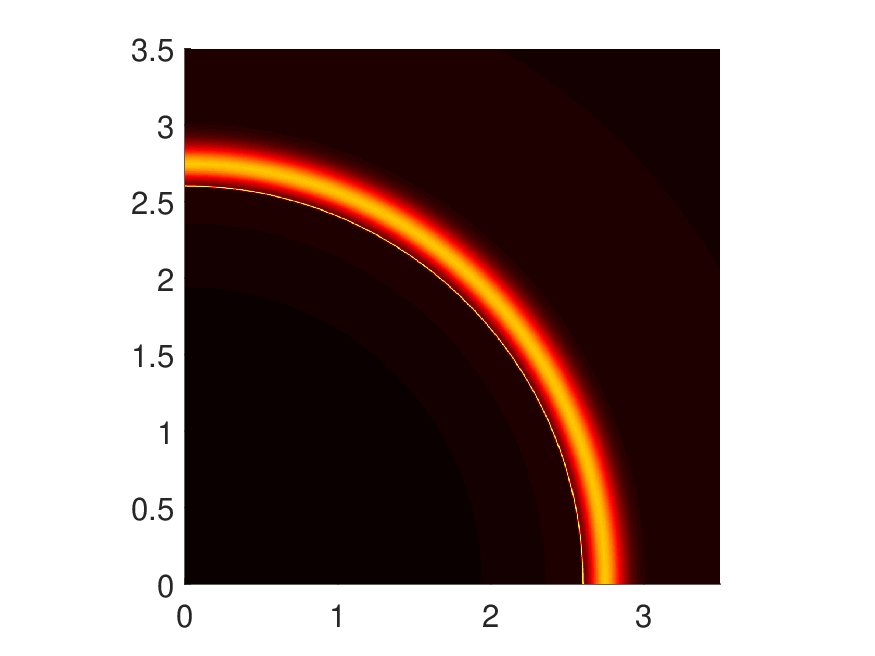}\hspace{-0.05\textwidth}\includegraphics[width=0.33\textwidth]{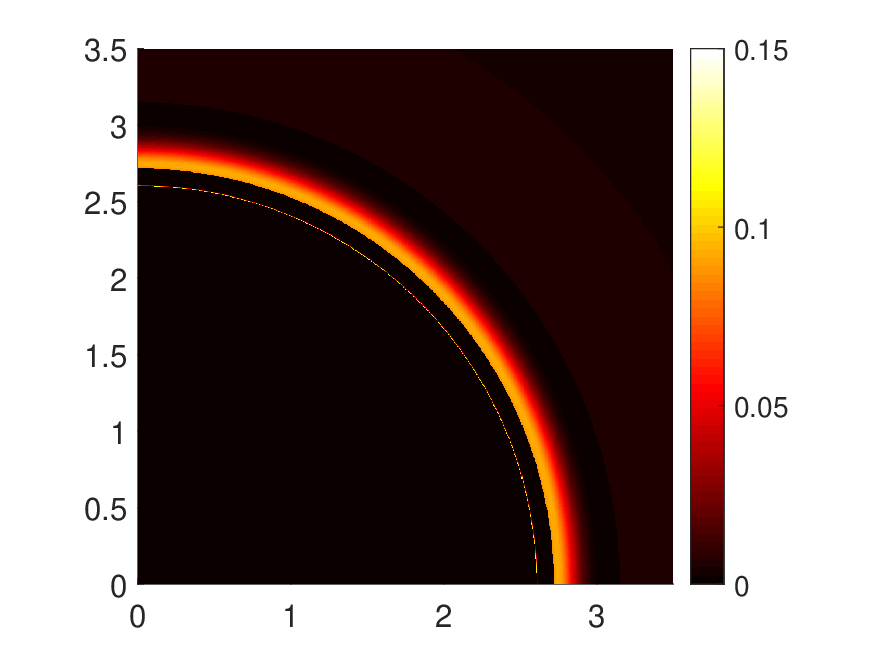}\\
\includegraphics[width=0.33\textwidth]{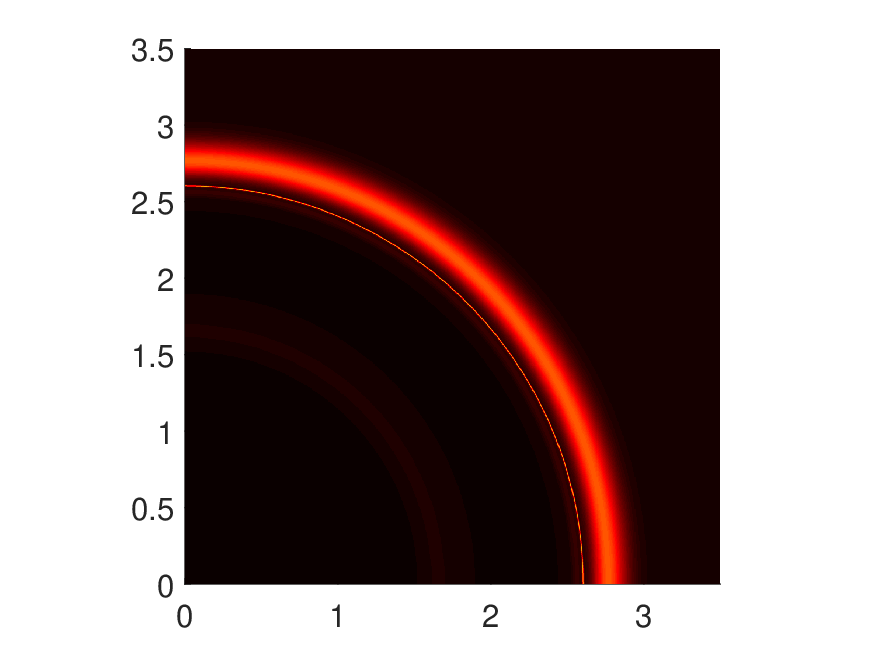}\hspace{-0.06\textwidth}\includegraphics[width=0.33\textwidth]{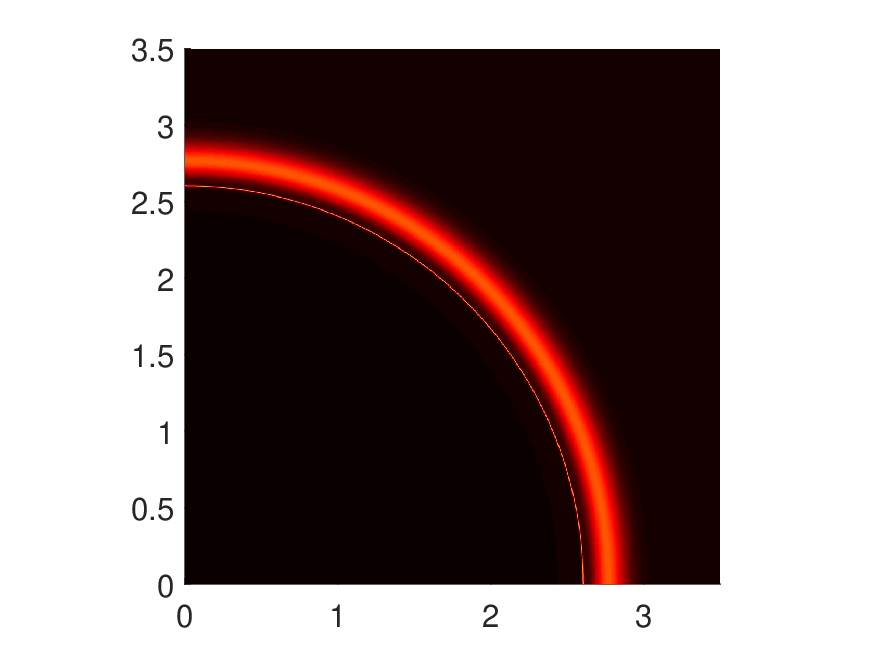}\hspace{-0.05\textwidth}\includegraphics[width=0.33\textwidth]{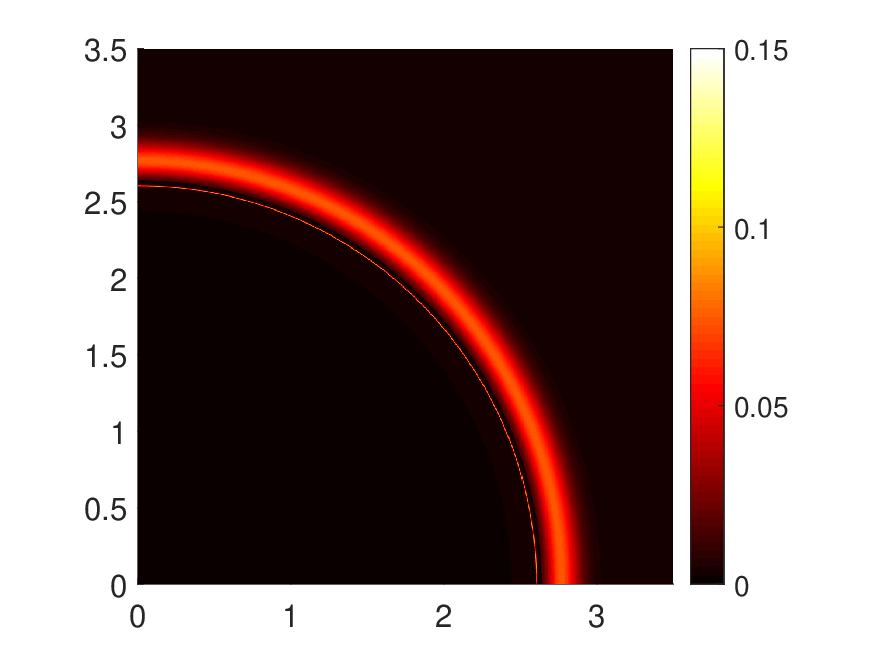}\\
\caption{Quarters of intensity images in the thin-disk accretion model. Same settings as in Fig. \ref{fig-IS}.}\label{fig-ID}
\end{figure}

For our thin-disk accretion model, the simulated intensity profiles and intensity images are presented in Figs. \ref{fig-PD} and \ref{fig-ID} respectively, where the notations and settings are the same as in Figs. \ref{fig-PS} and \ref{fig-IS}. Especially, the inner radius of the accretion disk $\rin/M=\rh/M=2$ in left panels, $\rin/M=\rph/M=3$ in the middle column, and $\rin/M=\risco/M=6,5.37,3.69$ in right panels.

Unlike Ref. \cite{Gralla:2019xty}, here the direct emission is not dominant in any panel. We have numerically investigated the three differences mentioned above: the redshift factor, the factor $D/(b\sqrt{A})$ and the emission coefficient. We find the redshift factor is responsible for the change. This is reasonable. Keep in mind that all photons are emitted from the gas on the thin disk. Because of the free fall of the accreting gas, the outer part emissions of both the photon ring and the lensing ring are boosted at $m\geq2$ (i.e., $\phi=3\pi/2,5\pi/2,\cdots$), their inner part emissions are deboosted at $m\geq2$, and all emissions are deboosted at $m=1$ (i.e., $\phi=\pi/2$), see the right panel of Fig. \ref{fig-traj}. As a result, the intensities of the photon ring and the lensing ring are partly enhanced, but all direct emissions are reduced in intensity. Indeed, as one can see in every panel of Fig. \ref{fig-ID}, there is a very thin photon ring located at $b=\bph=3\sqrt{3}\rh/2$ and a thicker lensing ring slightly outside it, but it is very hard to perceive the direct image.

From Figs. \ref{fig-PD} and \ref{fig-ID}, we can find out the trace of the length parameter $L$. As the ratio $L/\rh$ increases, the lensing ring moves farther away from the photon ring if $\rin=\rh$ or $\rin=\rph$. In contrast, if $\rin=\risco$, the two rings get closer to each other as $L/\rh$ increases.

For $\rin=\rh$ and $L/\rh=1.4$, in the left panel of Fig. \ref{fig-PD}, there is a bump in the blue dotted curve, which corresponds to a faint red band in the bottom-left panel of Fig. \ref{fig-ID}. The bump is located at $1.5<b/\rh<2$. It is the result of the inner portion of direct emissions with $b<b_m^{-}\approx2.51\rh$ \cite{Gralla:2019xty}. Let us take a closer look. In terms of $u=\rh/r$, Eq. \eqref{orbitr} can be rewritten as
\begin{equation}\label{orbitu}
\frac{du}{d\phi}=\mp\sqrt{\frac{\rh^2}{b^2}-u^2+u^3}.
\end{equation}
Paying attention to the region $b>\rh$ and neglecting temporarily the $u^3$ term on the right hand side, we integrate the equation to $\phi=\pi/2$ backwardly. This yields roughly $r_1(b)\sim b$. Then we can replace $u^3$ with $u^2\rh/b$ and get a better approximation
\begin{equation}\label{r1}
r_1(b)\sim\frac{b\sqrt{1-\frac{\rh}{b}}}{\sin\left(\frac{\pi}{2}\sqrt{1-\frac{\rh}{b}}\right)}.
\end{equation}
For $r_1(b)/\rh>\rin/\rh=1$, $1.5$, $1.84$, this gives $b>1.4$, $1.9$, $2.3$ approximately, in reasonable agreement with Fig. \ref{fig-PD}. As clear in Fig. \ref{fig-traj}, any direct emission is leaving the BH. Therefore, its contribution to the intensity \eqref{Isum} is
\begin{equation}\label{I1}
I_1(b)=\left.g_{+}^4j(r)\frac{D(r)}{b\sqrt{A(r)}}\right|_{r=r_1(b)}.
\end{equation}
We have inserted Eq. \eqref{r1} and $L/\rh=0$, $1$ and $1.4$ into this expression, and plotted it as a function of $b$. Indeed there is a bump around $b=1.6$ if and only if $L/\rh=1.4$.

\subsection{Comparison with observations}\label{subsect-obs}
Although the benchmark models studied in Secs. \ref{subsect-sph} and \ref{subsect-disk} are toy models, it is still possible to compare them against observations of the Event Horizon Telescope \cite{EventHorizonTelescope:2019dse,EventHorizonTelescope:2022xnr}. In accordance with Refs. \cite{EventHorizonTelescope:2019ggy,EventHorizonTelescope:2022xqj}, we define the fractional deviation parameter of the BH shadow size between the Event Horizon Telescope results and the model predictions as
\begin{equation}\label{delta}
\delta=\frac{d_{\mathrm{sh,EHT}}}{d_{\mathrm{sh,model}}}-1.
\end{equation}
In this expression, $d_{\mathrm{sh,EHT}}$ is the angular shadow diameter of M87* or Sgr A* measured by the Event Horizon Telescope, whereas $d_{\mathrm{sh,model}}$ is the angular shadow diameter predicted by our models using the following angular gravitational radius $\agr$: for M87*, $\agr$ is computed either from gas dynamics \cite{Walsh:2013uua} or from stellar dynamics \cite{Gebhardt:2011arl}; for Sgr A*, $\agr$ is computed from monitoring stellar orbits measured by VLTI \cite{GRAVITY:2021xju} and Keck \cite{Do:2019txf}. These observational results are summarized in Table \ref{tab-me}.

\begin{table}[ht]
\caption{Measured Parameters of M87* and Sgr A*.}\label{tab-me}
\begin{tabular*}{0.7\textwidth}{@{\extracolsep{\fill}}clcl}\hline
Source Name & Parameter & Best-fit Value & Measurement Method \\ \hline
M87* & Angular shadow diameter, $d_{\mathrm{sh,EHT}}$ & $42~\muas$ & EHT \cite{EventHorizonTelescope:2019dse}\\
~ & Angular gravitational radius, $\agr$ & $2.05~\muas$ & Gas dynamics \cite{Walsh:2013uua,EventHorizonTelescope:2019ggy}\\
~ & Angular gravitational radius, $\agr$ & $3.62~\muas$ & Stellar dynamics \cite{Gebhardt:2011arl,EventHorizonTelescope:2019ggy}\\ \hline
Sgr A* & Angular shadow diameter, $d_{\mathrm{sh,EHT}}$ & $48.7~\muas$ & EHT \cite{EventHorizonTelescope:2022xnr}\\
~ & Angular gravitational radius, $\agr$ & $5.125~\muas$ & Stellar orbits (VLTI) \cite{GRAVITY:2021xju,EventHorizonTelescope:2022xnr}\\
~ & Angular gravitational radius, $\agr$ & $4.92~\muas$ & Stellar orbits (Keck) \cite{Do:2019txf,EventHorizonTelescope:2022xnr}\\
\hline\end{tabular*}
\end{table}
\begin{table}[ht]
\caption{Deviation Parameters of M87* and Sgr A*.}\label{tab-de}
\begin{tabular*}{0.7\textwidth}{@{\extracolsep{\fill}}clcc}\hline
Source Name & Measurement Method & $\delta$ in two special models & $\delta$ in other benchmark models \\ \hline
M87* & Gas dynamics \cite{Walsh:2013uua,EventHorizonTelescope:2019ggy} & $0.90$ & $0.97$\\
~ & Stellar dynamics \cite{Gebhardt:2011arl,EventHorizonTelescope:2019ggy} & $0.074$ & $0.12$\\ \hline
Sgr A* & Stellar orbits (VLTI) \cite{GRAVITY:2021xju,EventHorizonTelescope:2022xnr} & $-0.12$ & $-0.086$\\
~ & Stellar orbits (Keck) \cite{Do:2019txf,EventHorizonTelescope:2022xnr} & $-0.083$ & $-0.048$\\
\hline\end{tabular*}
\end{table}

The angular gravitational radius is defined by
\begin{equation}\label{agr}
\agr=\frac{GM}{Dc^2}=\frac{\rh}{2D}
\end{equation}
with $D$ being the BH distance, while the angular shadow diameter is defined as $d_{\mathrm{sh}}=2\rsh/D$ with $\rsh$ being the radius of shadow boundary predicted by our models. From Figs. \ref{fig-PS}, \ref{fig-IS}, \ref{fig-PD} and \ref{fig-ID}, we can see the boundary of BH shadow is located at $\rsh\simeq2.7\rh$ for two special thin-disk accretion models: $\rin=\risco$ with $L/\rh=0$ or $1$. Inserting Eq. \eqref{agr} and this value of $\rsh$ into the definition of angular shadow diameter, we can predict $d_{\mathrm{sh,model}}\simeq10.8\agr$. For other benchmark models, the radius of shadow boundary is $\rsh\simeq2.6\rh$, which implies $d_{\mathrm{sh,model}}\simeq10.4\agr$. Making use of the measured parameters in Table \ref{tab-me}, we can then calculate the fractional deviation parameter $\delta$. The results are relegated in Table \ref{tab-de}. For M87*, the two special models fit better than other benchmark models. However, for Sgr A*, the two special models fit worse than other benchmark models. Barring measurement uncertainties, the deviations can be attributed to the BH spins, which are beyond the scope of this paper.

\section{Conclusion}\label{sect-con}
In this paper, we have polished two analytical toy models of BH accretions and applied them to images of nonsingular nonrotating BHs in conformal gravity, or namely the conformal Schwarzschild BHs. A relativistic expression of the emission coefficient per unit solid angle \eqref{j} has been developed by assuming that the emission energy comes exactly from the gravitational potential energy in static spherical spacetimes. A new expression of the observed image intensity in the sum form \eqref{Isum} has been derived rigorously for static spherical BHs surrounded by thickless accretion disks. Considering a nonsingular nonrotating BH surrounded by a free-falling accreting gas, we simulated its intensity profiles and shadow images for various values of length parameter in the conformal scale factor and for various radii of inner accretion boundary.

We have proved that the shadow radius of the conformal Schwarzschild BH is determined solely by its mass. With their metric given by Eqs. \eqref{metric-conSch} and \eqref{S}, the conformal Schwarzschild BH and the Schwarzschild BH of the same mass have the same coordinate radius of photon sphere as well as the same shadow radius. However, their shadow images are different in intensity distribution. In the spherical accretion model, the conformal scale factor affects the contrast between inside and outside of the shadow. In the thin-disk accretion model, it influences the width of a gap between the lensing ring and the photon ring. Unfortunately, in both models, the effects of the conformal factor are degenerate with the influences from the radius of the inner accretion boundary.

Our investigations have an interesting implication even restricted to Schwarzschild BHs. In contrast with a gas at rest in Ref. \cite{Gralla:2019xty}, the free fall of the accreting gas in our paper dims the emission inside the photon sphere and, in the thin disk model, brightens the photon-ring and lensing-ring emissions. As a result, the shadow boundary is located at or slightly outside the photon sphere irrespective of the radius of the inner accretion boundary. As we have discussed in Sec. \ref{sect-image}, this is most likely attributed to the trajectory-dependent Doppler boosting and deboosting effects. It deserves an exhaustive research elsewhere to confirm the generality and the cause of this result.

\begin{acknowledgments}
This work is supported by National Science Foundation of China grant No.~11105091. We thank Yanni Zhu for useful discussions.
\end{acknowledgments}


\begin{thebibliography}{99}
\bibitem{Synge:1966okc}
J.~L.~Synge,
Mon. Not. Roy. Astron. Soc. \textbf{131}, no.3, 463-466 (1966)

\bibitem{Bardeen:1973}
J.~M.~Bardeen,
Timelike and null geodesics of the Kerr metric,
Gordon Breach, Science Publishers, New York (1973)

\bibitem{Luminet:1979nyg}
J.~P.~Luminet,
Astron. Astrophys. \textbf{75}, 228-235 (1979)

\bibitem{Falcke:1999pj}
H.~Falcke, F.~Melia and E.~Agol,
Astrophys. J. Lett. \textbf{528}, L13 (2000)
[arXiv:astro-ph/9912263 [astro-ph]].

\bibitem{EventHorizonTelescope:2019dse}
K.~Akiyama \textit{et al.} [Event Horizon Telescope],
Astrophys. J. Lett. \textbf{875}, L1 (2019)
[arXiv:1906.11238 [astro-ph.GA]].

\bibitem{EventHorizonTelescope:2022xnr}
K.~Akiyama \textit{et al.} [Event Horizon Telescope],
Astrophys. J. Lett. \textbf{930}, no.2, L12 (2022)

\bibitem{Israel:1967wq}
W.~Israel,
Phys. Rev. \textbf{164}, 1776-1779 (1967)

\bibitem{Robinson:1975bv}
D.~C.~Robinson,
Phys. Rev. Lett. \textbf{34}, 905-906 (1975)

\bibitem{Psaltis:2018xkc}
D.~Psaltis,
Gen. Rel. Grav. \textbf{51}, no.10, 137 (2019)
[arXiv:1806.09740 [astro-ph.HE]].

\bibitem{Bambi:2019tjh}
C.~Bambi, K.~Freese, S.~Vagnozzi and L.~Visinelli,
Phys. Rev. D \textbf{100}, no.4, 044057 (2019)
[arXiv:1904.12983 [gr-qc]].

\bibitem{Vagnozzi:2019apd}
S.~Vagnozzi and L.~Visinelli,
Phys. Rev. D \textbf{100}, no.2, 024020 (2019)
[arXiv:1905.12421 [gr-qc]].

\bibitem{Allahyari:2019jqz}
A.~Allahyari, M.~Khodadi, S.~Vagnozzi and D.~F.~Mota,
JCAP \textbf{02}, 003 (2020)
[arXiv:1912.08231 [gr-qc]].

\bibitem{Khodadi:2020jij}
M.~Khodadi, A.~Allahyari, S.~Vagnozzi and D.~F.~Mota,
JCAP \textbf{09}, 026 (2020)
[arXiv:2005.05992 [gr-qc]].

\bibitem{Ghosh:2022gka}
R.~Ghosh, M.~Rahman and A.~K.~Mishra,
Eur. Phys. J. C \textbf{83}, no.1, 91 (2023)
[arXiv:2209.12291 [gr-qc]].

\bibitem{Perlick:2021aok}
V.~Perlick and O.~Y.~Tsupko,
Phys. Rept. \textbf{947}, 1-39 (2022)
[arXiv:2105.07101 [gr-qc]].

\bibitem{Bronzwaer:2021lzo}
T.~Bronzwaer and H.~Falcke,
Astrophys. J. \textbf{920}, no.2, 155 (2021)
[arXiv:2108.03966 [astro-ph.HE]].

\bibitem{Wang:2022kvg}
M.~Wang, S.~Chen and J.~Jing,
Commun. Theor. Phys. \textbf{74}, no.9, 097401 (2022)
[arXiv:2205.05855 [gr-qc]].

\bibitem{Vagnozzi:2022moj}
S.~Vagnozzi, R.~Roy, Y.~D.~Tsai, L.~Visinelli, M.~Afrin, A.~Allahyari, P.~Bambhaniya, D.~Dey, S.~G.~Ghosh and P.~S.~Joshi, \textit{et al.}
Class. Quant. Grav. \textbf{40}, no.16, 165007 (2023)
[arXiv:2205.07787 [gr-qc]].

\bibitem{Chen:2022scf}
S.~Chen, J.~Jing, W.~L.~Qian and B.~Wang,
Sci. China Phys. Mech. Astron. \textbf{66}, no.6, 260401 (2023)
[arXiv:2301.00113 [astro-ph.HE]].

\bibitem{Olivares:2018abq}
H.~Olivares, Z.~Younsi, C.~M.~Fromm, M.~De Laurentis, O.~Porth, Y.~Mizuno, H.~Falcke, M.~Kramer and L.~Rezzolla,
Mon. Not. Roy. Astron. Soc. \textbf{497}, no.1, 521-535 (2020)
[arXiv:1809.08682 [gr-qc]].

\bibitem{Herdeiro:2021lwl}
C.~A.~R.~Herdeiro, A.~M.~Pombo, E.~Radu, P.~V.~P.~Cunha and N.~Sanchis-Gual,
JCAP \textbf{04}, 051 (2021)
[arXiv:2102.01703 [gr-qc]].

\bibitem{Modesto:2016max}
L.~Modesto and L.~Rachwal,
[arXiv:1605.04173 [hep-th]].

\bibitem{Bambi:2016wdn}
C.~Bambi, L.~Modesto and L.~Rachwa\l{},
JCAP \textbf{05}, 003 (2017)
[arXiv:1611.00865 [gr-qc]].

\bibitem{Bambi:2017yoz}
C.~Bambi, Z.~Cao and L.~Modesto,
Phys. Rev. D \textbf{95}, no.6, 064006 (2017)
[arXiv:1701.00226 [gr-qc]].

\bibitem{Narayan:2019imo}
R.~Narayan, M.~D.~Johnson and C.~F.~Gammie,
Astrophys. J. Lett. \textbf{885}, no.2, L33 (2019)
[arXiv:1910.02957 [astro-ph.HE]].

\bibitem{Gralla:2019xty}
S.~E.~Gralla, D.~E.~Holz and R.~M.~Wald,
Phys. Rev. D \textbf{100}, no.2, 024018 (2019)
[arXiv:1906.00873 [astro-ph.HE]].

\bibitem{Zhu:2021tgb}
Y.~Zhu and T.~Wang,
Phys. Rev. D \textbf{104}, no.10, 104052 (2021)
[arXiv:2109.08463 [gr-qc]].

\bibitem{Mannheim:1991ez}
P.~D.~Mannheim,
Gen. Rel. Grav. \textbf{25}, 697-715 (1993)

\bibitem{Hobson:2022ahe}
M.~Hobson and A.~Lasenby,
Eur. Phys. J. C \textbf{82}, no.7, 585 (2022)
[arXiv:2206.08097 [gr-qc]].

\bibitem{Mannheim:2021fql}
P.~D.~Mannheim,
Class. Quant. Grav. \textbf{39}, no.24, 245001 (2022)
[arXiv:2105.08556 [gr-qc]].

\bibitem{Song:2021ziq}
Y.~Song,
Eur. Phys. J. C \textbf{81}, no.10, 875 (2021)
[arXiv:2108.00696 [gr-qc]].

\bibitem{Carroll:2004st}
S.~M.~Carroll,
``Spacetime and Geometry: An Introduction to General Relativity,'' Cambridge University Press, 2019.

\bibitem{Zeng:2020vsj}
X.~X.~Zeng and H.~Q.~Zhang,
Eur. Phys. J. C \textbf{80}, no.11, 1058 (2020)
[arXiv:2007.06333 [gr-qc]].

\bibitem{Peng:2020wun}
J.~Peng, M.~Guo and X.~H.~Feng,
Chin. Phys. C \textbf{45}, no.8, 085103 (2021)
[arXiv:2008.00657 [gr-qc]].

\bibitem{Li:2021ypw}
G.~P.~Li and K.~J.~He,
Eur. Phys. J. C \textbf{81}, no.11, 1018 (2021)

\bibitem{Bambi:2013nla}
C.~Bambi,
Phys. Rev. D \textbf{87}, 107501 (2013)
[arXiv:1304.5691 [gr-qc]].

\bibitem{Bisnovatyi-Kogan:2022ujt}
G.~S.~Bisnovatyi-Kogan and O.~Y.~Tsupko,
Phys. Rev. D \textbf{105}, no.6, 064040 (2022)
[arXiv:2201.01716 [gr-qc]].

\bibitem{EventHorizonTelescope:2019ggy}
K.~Akiyama \textit{et al.} [Event Horizon Telescope],
Astrophys. J. Lett. \textbf{875}, no.1, L6 (2019)
[arXiv:1906.11243 [astro-ph.GA]].

\bibitem{EventHorizonTelescope:2022xqj}
K.~Akiyama \textit{et al.} [Event Horizon Telescope],
Astrophys. J. Lett. \textbf{930}, no.2, L17 (2022)

\bibitem{Walsh:2013uua}
J.~L.~Walsh, A.~J.~Barth, L.~C.~Ho and M.~Sarzi,
Astrophys. J. \textbf{770}, 86 (2013)
[arXiv:1304.7273 [astro-ph.CO]].

\bibitem{Gebhardt:2011arl}
K.~Gebhardt, J.~Adams, D.~Richstone, T.~R.~Lauer, S.~M.~Faber, K.~Gultekin, J.~Murphy and S.~Tremaine,
Astrophys. J. \textbf{729}, 119 (2011)
[arXiv:1101.1954 [astro-ph.CO]].

\bibitem{GRAVITY:2021xju}
R.~Abuter \textit{et al.} [GRAVITY],
Astron. Astrophys. \textbf{657}, L12 (2022)
[arXiv:2112.07478 [astro-ph.GA]].

\bibitem{Do:2019txf}
T.~Do, A.~Hees, A.~Ghez, G.~D.~Martinez, D.~S.~Chu, S.~Jia, S.~Sakai, J.~R.~Lu, A.~K.~Gautam and K.~K.~O'Neil, \textit{et al.}
Science \textbf{365}, no.6454, 664-668 (2019)
[arXiv:1907.10731 [astro-ph.GA]].


\end{thebibliography}
\end{document}